\documentclass[conference]{IEEEtran}

\usepackage{cite}
\usepackage[cmex10]{amsmath}
\usepackage{amsfonts,amssymb}
\usepackage[pdftex]{graphicx}

\begin{document}

\title{The Capacity of Adaptive Group Testing}

\author{\IEEEauthorblockN{Leonardo Baldassini}\IEEEauthorblockA{School of Mathematics\\ University of Bristol\\ University Walk, Bristol\\ BS8 1TW, UK \\ Email: Leonardo.Baldassini@bristol.ac.uk}
\and
\IEEEauthorblockN{Oliver Johnson} \IEEEauthorblockA{School of Mathematics\\ University of Bristol\\ University Walk, Bristol\\ BS8 1TW, UK \\ Email: O.Johnson@bristol.ac.uk}
\and
\IEEEauthorblockN{Matthew Aldridge} \IEEEauthorblockA{Heilbronn Institute for Mathematical Research\\ School of Mathematics\\ University of Bristol\\ University Walk, Bristol, BS8 1TW, UK \\ Email: M.Aldridge@bristol.ac.uk}
}

\newcommand{\vc}[1]{{\mathbf{ #1}}}
\newcommand{\var}{{\rm{Var\;}}}
\newcommand{\cov}{{\rm{Cov\;}}}
\newcommand{\tends}{\rightarrow \infty}
\newcommand{\ep}{{\mathbb {E}}}
\newcommand{\pr}{{\mathbb {P}}}
\newcommand{\re}{{\mathbb {R}}}
\newcommand{\I}{\mathbb {I}}
\newcommand{\Z}{{\mathbb {Z}}}
\newcommand{\blah}[1]{}
\newcommand{\pin}{ p_i(\mbox{new})}
\newcommand{\pio}{ p_i(\mbox{old})}
\newcommand{\suc}{{\rm suc}}
\newcommand{\SSS}{{\mathcal S}}
\newcommand{\AAA}[1]{{\mathcal A}_{\vc{#1}}}
\newcommand{\AAS}[1]{A_{\vc{#1}}}
\newcommand{\II}{{\mathbb I}}
\newcommand{\defset}{{\mathcal D}}

\newcommand{\given}{\mid}

\newcommand{\idx}[3] {\ensuremath{{#1}^{(#2)}_{#3}}}
\newcommand{\df}[2]{\displaystyle{\frac{#1}{#2}}}

\newtheorem{theorem}{Theorem}[section]
\newtheorem{lemma}[theorem]{Lemma}
\newtheorem{proposition}[theorem]{Proposition}
\newtheorem{corollary}[theorem]{Corollary}
\newtheorem{conjecture}[theorem]{Conjecture}
\newtheorem{definition}[theorem]{Definition}
\newtheorem{example}[theorem]{Example}
\newtheorem{condition}{Condition}
\newtheorem{main}{Theorem}
\newtheorem{remark}[theorem]{Remark}

\maketitle

\begin{abstract}
We define capacity for group testing problems and deduce bounds for the capacity of a variety of noisy models, based on the capacity of equivalent noisy communication channels. For noiseless adaptive group testing we prove
 an information-theoretic
lower bound which tightens a bound of Chan et al. This can be combined with a
performance analysis of a version of Hwang's adaptive group testing algorithm, in order to
deduce the capacity of noiseless and erasure group testing models.
\end{abstract}

\IEEEpeerreviewmaketitle

\section{Introduction and notation}

We consider noiseless group testing, as introduced by
Dorfman \cite{dorfman} in the context of testing populations of soldiers for syphilis.
Group testing has recently been used to screen DNA samples for rare
alleles by a pooling strategy, and relates to compressive sensing  -- see \cite{erlich, shental, aksoylar, malyutov}.
It also offers advantages for spectrum sharing in cognitive radio models \cite{atia2}.

Suppose we have a population of $N$ items, $K$ of which are
defective. At the $i$th stage, we pick a subset $\SSS_i$ of the population, and test all the items in $\SSS_i$ together.
If $\SSS_i$ contains at least one defective item, the test is positive. If $\SSS_i$ contains no defective item, the test is negative.

The group testing problem requires us to 
infer which items are defective, using the smallest  number
of tests. We design the tests to minimise the
(expected) number of tests required. For simplicity, this paper assumes the number of defectives $K$ is known.

The problem as described above
 is referred to as noiseless, since the test output 
is a known deterministic function of the items tested. In this case,
positive and negative tests allow complementary
inferences. If the $i$th test is negative, all the items in $\SSS_i$
 are not defective, so we need not test them again.
If the $i$th test is positive,  at least one item in 
$\SSS_i$ is
defective, but it will usually require further testing to discover which.

It is natural to consider noisy models of group testing. For example, \cite{atia} considers
an ``additive model" where a negative test may erroneously be 
reported positive with probability  $p$. The paper \cite{chan} considers a ``symmetric error model"
where the test outputs are transmitted through memoryless binary symmetric channels with
error probability $p$. Finally, we consider an ``erasure model", where with probability $p$,
the test fails, and returns an erasure symbol.

In this paper we consider adaptive testing, where the test design depends 
on the outcome of previous tests, in contrast to
the non-adaptive case, where the entire sequence of tests is specified in
advance.
In other contexts (see for example \cite{haupt}), the ability to perform
adaptive measurements can provide significant improvements (for example,
moving problems from polynomial accuracy to exponential accuracy). 
Perhaps surprisingly (see \cite{aldridge}), in the context of group testing,
the improvements are smaller, in that allowing adaptive algorithms only
provides an improvement by a constant factor in the number of tests bounds (see
for example Theorems \ref{thm:rbt} and \ref{thm:non-adaptive}).

In the 1970s, Malyutov and co-authors studied group testing as a channel coding problem -- see Malyutov's survey paper \cite{malyutov} for a historical perspective on this and other approaches. In this work we will follow the notation of 
Atia and Saligrama \cite{atia} who
further studied group testing as a channel coding problem. We make analogies between the number of tests $T$
required and the code block length, and between the number of possible
defective sets $\binom{N}{K}$ and the number of possible messages. This suggests
that in the spirit of Shannon \cite{shannon2} we should consider the rate as
$ \log_2 \binom{N}{K}/T$ and introduce the capacity:

\begin{definition} \label{def:capacity}
  Consider a sequence of group testing problems, 
  indexed by the number of items $N = 1, 2, \ldots$. The $N$th
  problem has $K = K(N)$ defective items, where $1 \leq K \leq N$.
  and $T=T(N)$ tests are available. We refer to a constant $C$ as the group testing capacity if for any $\epsilon > 0$:
  \begin{enumerate}
    \item any sequence of algorithms with
      \begin{equation} \label{eq:lower}
        \liminf_{N \tends} \frac{\log_2 \binom{N}{K(N)} }{T(N)} \geq C+ \epsilon,
      \end{equation}
      has success probability tending to 0,
    \item and there exists a sequence of algorithms with
      \begin{equation} \label{eq:upper}
        \liminf_{N \tends} \frac{\log_2 \binom{N}{K(N)}}{T(N)}  \geq C - \epsilon
      \end{equation}
      with success probability tending to 1.
  \end{enumerate}
\end{definition}

Malyutov \cite[Section 2]{malyutov} gives a different definition of 
capacity, considering the rate (in our notation) as
$(\log N)/T$ for given $K$.

Group testing makes particular
gains when the set of defectives is sparse, in that $K = o(N)$.
Otherwise, we could be accurate to within a multiple of the optimal
number of tests simply by testing all items individually.
For benchmarking
purposes, we use the parameterisation of 
\cite{haupt}, \cite{donoho} and others, taking $K = N^{1-\beta}$ for some $0 < \beta < 1$.
The main result of the paper is the following:

\begin{theorem} \label{thm:main}
  The capacity of the adaptive noiseless group testing problem
  is  $C = 1$,
  in any regime such that $K=o(N)$.
\end{theorem}

Theorem \ref{thm:main} is proved in Section \ref{sec:lower}  by combining
existing performance guarantees  (summarised in Section \ref{sec:existing}) with a new information theoretic lower
bound, Theorem \ref{thm:basic}.

It is natural to consider the capacity of noisy group testing problems under different noise models. The sequence of outcomes of tests can be considered as a binary codeword encoding the defective set. However, tests can only verify whether a subset intersects the defective set, whereas in a general communication channel the encoding being agreed by the receiver and the sender may incorporate more information. Hence we can state a new key principle:
\begin{quote} 
if it exists, the capacity of a noisy group testing problem can never exceed the capacity of the equivalent communication channel.
\end{quote}
We then state the following bounds on group testing capacities:
\begin{theorem} \label{thm:noisy} 
\begin{enumerate}
\item \label{it:erasure} In any regime with $K = o(N)$, for the ``erasure model'' with probability $p$, $C = 1-p$.
\item For the ``symmetric error model" with probability $p$ of \cite{chan}, if it exists $C \leq 1-h(p)$,  
 with $h$ the binary entropy. It is natural to conjecture  $C=1-h(p)$. 
\item For the ``additive model'' with probability $p$ of \cite{atia}, if it exists
$C \leq \log(1+(1-p)p^{p/(1-p)})$,
 the capacity of a Z-channel. Again we conjecture equality holds.
\end{enumerate}
\end{theorem}
\begin{IEEEproof} In each case, the upper bound follows from the key principle above, using standard values of
channel capacities and the strong converse to Shannon's channel coding theorem \cite{wolfowitz}. The exact value for \ref{it:erasure}) is derived in Section \ref{sec:lower} below. \end{IEEEproof}
The structure of the remainder of the paper is as follows.  In Section
\ref{sec:existing} we review existing approaches to the group testing 
problem, describing upper bounds   on the number of the tests required, including Hwang's
Generalized Binary Splitting Algorithm.
In Section \ref{sec:lower} we prove lower bounds on the number of tests required, giving an upper bound
on the success probability, and proving the capacity theorem, Theorem \ref{thm:main}.
In Section \ref{sec:ubound} we give simulation evidence which shows that these bounds are tight for
a range of success probabilities, and prove Theorem
\ref{thm:ubound} which shows we can improve on Hwang's algorithm, and
can be adapted to suggest bounds on success probabilities in general.
\section{Existing performance guarantees} \label{sec:existing}
We first describe some simple upper bounds on the number of tests required by group testing
algorithms, though in Figure \ref{fig:hist} such bounds  are
plotted in terms of success probabilities. Clearly, an upper bound on the number of tests required
relates to a lower bound on the success probability.

Early approaches to the group testing problem involved deterministic
(combinatorial)
test designs, as reviewed in \cite{du}. In particular, at least
$\Omega( K^2 \log N/\log K)$ tests are required for such designs in the
non-adaptive case to guarantee success.

More recent work has focussed on randomised test designs. Initial bounds on sample complexity using information-theoretic methods are surveyed in \cite{malyutov1}. Atia and Saligrama \cite{atia}
were able to show that $O( K \log N)$ tests would suffice with
high probability, for non-adaptive group testing, even in the noisy case.
This work was developed by Chan et al \cite{chan}, who provided explicit
algorithms, and corresponding bounds on their performance, in the non-adaptive
case. More recently, Aksoylar et al. \cite{aksoylar} have unified this and other related problems in sparse signal recovery in Shannon-theoretic terms, hence producing sample-size bounds and recovery conditions which hold in general. Following an approach inspired by LDPC codes, in \cite{wadayama} Wadayama proposes an analysis of non-adaptive group testing based on sparse bipartite graphs in the regime where $K$ asymptotically scales as $pN$. In a paper \cite{johnson33} in preparation
we strengthen some of the results of \cite{chan}.

 Performance of many adaptive algorithms is guaranteed by the
following simple lemma:


\begin{lemma}\label{lem:binsearch}
Given a set of $b$ items that is known to contain at least one defective item, label its elements ${1,\ldots,b}$ and let $L$ be the smallest label of a defective item in the set. Then 
in $\lceil\log_2 b\rceil$ tests we can discover with certainty that item $L$
is defective and that  items $1,\ldots, L-1$ are all non-defective. 
\end{lemma}

\begin{IEEEproof}
Use the following recursive procedure, which we refer to as `binary search'. If necessary,
we add `dummy items' to create a set of size $2^{\lceil \log_2 b \rceil}$.
At each stage, given a set of size $S$ which is guaranteed to contain a defective,
we label its items with integers $\{ 1, 2, \ldots, S \}$. We test the items with
labels $\{ 1, 2, \ldots,  S/2  \}$.
\begin{enumerate}
  \item If the test is positive, we have a set of  half the previous size,
    guaranteed to contain a defective.
  \item If the test is negative, we know that items $\{S/2 +1, \ldots, S\}$
    must contain a defective item.
\end{enumerate}
Each test therefore  halves the size of the set, with it remaining
guaranteed to contain a defective item. The property of finding the defective with the smallest label follows easily by induction and by the fact that at each step we always test items $\{1, 2, \ldots, S/2\}$ before $\{S/2+1,\ldots, S\}$, and discard the latter if the former is found to be positive.
\end{IEEEproof}


A simple adaptive
algorithm with guaranteed performance bounds is given by 
Repeated Binary Testing   \cite[p24--5]{du}.
The algorithm simply performs binary search on the set of size
$N$ to find a defective item.
This item is then removed
from consideration, and the next round of testing carries out binary
testing on a set of size $N-1$. Repeatedly using Lemma \ref{lem:binsearch},
it is clear that this algorithm provides
a performance guarantee for  adaptive testing.

\begin{theorem}[\cite{du}] \label{thm:rbt}
  Repeated Binary Testing is guaranteed to succeed in
  $K \lceil \log_2 N \rceil \leq K \log_2 N + K$ tests.
\end{theorem}

The Repeated Binary Testing algorithm is inefficient, in that
each binary search starts by testing large sets, which are
very likely to contain at least one defective. In that sense,
the early tests in each round are very uninformative.

Hwang's
Generalized Binary Splitting  Algorithm (HGBSA) \cite{hwang} is designed to overcome this
problem.
Hwang suggests testing groups of size $2^{\alpha}$, where
$\alpha$ is an integer chosen to ensure that the probability of
the test being positive is close to $1/2$. If the test is negative,
all the items in it can be immediately classified as non-defective.
If the test is positive, it must contain a defective, which can be found in
$\alpha$ tests using the binary search procedure of Lemma \ref{lem:binsearch}
above.

Using this procedure, Hwang \cite[Theorem 1]{hwang} deduces an upper bound on the 
number of tests required, which further analysis (see \cite[Corollary 2.2.2]{du}) shows 
is close to optimal in the
sense discussed in the proof of Theorem \ref{thm:main}.

\begin{theorem}[\cite{du}] \label{thm:hwang}
  Given a problem with $K$ defectives in a population of size $N$,
  Hwang's adaptive Generalized Binary Splitting Algorithm is guaranteed
  to succeed using 
  \begin{equation} \label{eq:hwang}
    T = \log_2 \binom{N}{K} + K \mbox{ tests.}
  \end{equation}
\end{theorem}

In order to see the gain due to adaptivity, this bound can be
compared with the following result, Theorem 4 of \cite{chan}.

\begin{theorem}[\cite{chan}] \label{thm:non-adaptive}
  The Combinatorial Orthogonal Matching Pursuit (COMP) algorithm
  recovers the defective set with error probability $\leq N^{-\delta}$,
  using $T  = ( (1+\delta) e) K \ln N$ tests.
\end{theorem}

Theorems \ref{thm:hwang} and \ref{thm:non-adaptive} can be compared using
well-known bounds on the binomial coefficients, that is for all $K$ and $N$:
\begin{equation} \label{eq:binomlog}
  K \log_2(N/K) \leq \log_2 \binom{N}{K} \leq K \log_2 (N e/K),
\end{equation}
In the regime $K(N) = N^{1-\beta}$  considered by \cite{donoho,haupt}, we can see that 
(\ref{eq:binomlog}) and Theorem \ref{thm:hwang} can be combined to show
that Hwang's adaptive algorithm is guaranteed to succeed in
$$ \beta K \log_2 N + K (\log_2 e + 1) $$
tests. In contrast, Theorem \ref{thm:non-adaptive} shows that the (non-adaptive) COMP algorithm 
\cite{chan} requires
$$ \left( \frac{ (1+ \delta) e}{\log_2 e} \right) K \log_2 N \approx
1.88 (1+\delta) K \log_2 N,	$$
showing that adaptivity offers asymptotic gains over the 
bounds of Theorem \ref{thm:non-adaptive} which are greatest when
$\beta$ is small -- which is when there are relatively many defectives.
\section{New information-theoretic lower bound} \label{sec:lower}
We contrast the performance guarantees of Section \ref{sec:existing}
 with  a new  information-theoretic lower bound, Theorem \ref{thm:basic}, which applies to both adaptive and non-adaptive
group testing. It is possible that a tighter lower bound could be found
if we restrict ourselves to non-adaptive tests, but we are not aware of
such a result.

\begin{theorem} \label{thm:basic}
  Consider testing a set of $N$ items with $K$ 
  defectives. Any algorithm to recover the defective set with $T$ tests has success
  probability $\pr(\suc)$ satisfying
  \begin{equation}
    \pr( \suc ) \leq \frac{2^T}{\binom{N}{K}}.
    \label{eq:basic}
  \end{equation}
\end{theorem}
 
\begin{IEEEproof}
Given a population of $N$ objects, we write $\Sigma_{N,K}$ for the 
collection of subsets of size $K$ from the population. Further,
we write $\defset$ for the true defective set.

The testing procedure naturally 
defines a mapping $\theta: \Sigma_{N,K} \rightarrow \{ 0, 1 \}^T$.
That is, given a putative defective set $S \in \Sigma_{N,K}$,
write $\theta(S)$ to be the vector of test outcomes, with 
positive tests represented as 1s and negative ones represented as 0s.
For each vector $\vc{y} \in \{ 0,1 \}^T$, write $\AAA{y} \subseteq
\Sigma_{N,K}$ for the inverse image of $\vc{y}$ under $\theta$,
$$ \AAA{y} = \theta^{-1}(\vc{y}) = 
\left\{ S \in \Sigma_{N,K}: \theta( S) = \vc{y} \right \},$$
and write $\AAS{y} = | \AAA{y} |$ for the size of $\AAA{y}$. 

The role of an algorithm which decodes the outcome of the tests is to mimic the effect of the
inverse image map $\theta^{-1}$. Given a  test output  $\vc{y}$,
the optimal decoding algorithm would use a lookup table to find the
inverse image $\AAA{y}$. If this inverse image $\AAA{y} = \{ S \}$
has size $\AAS{y} =
1$, we can be certain that the defective set
was $S$. In general, if size $\AAS{y} \geq 1$, we cannot do 
better than to
pick uniformly among $\AAA{y}$, with success probability $1/\AAS{y}$.
(We can ignore empty $\AAA{y}$, since we are only concerned with
vectors $\vc{y}$ which occur as a test output).

Hence overall, the probability of recovering a defective set
$S$ is $1/|\mathcal A_{ \theta(S)}|$, depending only on $\theta(S)$. 
We can write the following expression for the success
probability, conditioning over all the equiprobable values of the 
defective set:
\begin{align*}
  \pr( \suc )  
        &= \sum_{S \in \Sigma_{N,K}} 
         \pr( \suc \given  \defset=S ) \frac{1}{\binom{N}{K}}  \\
    &= \frac{1}{\binom{N}{K}} \sum_{S \in \Sigma_{N,K}} 
         \sum_{\vc{y} \in \{0,1 \}^T} \II( \theta(S) = \vc{y})
         \pr( \suc \given \defset =S ) \\
    &= \frac{1}{\binom{N}{K}} \sum_{S \in \Sigma_{N,K}} 
         \sum_{\vc{y} \in \{0,1 \}^T: \AAS{y} \geq 1} 
         \II( \theta(S) = \vc{y}) \frac{1}{\AAS{y}} \\
    &= \frac{1}{\binom{N}{K}} \sum_{\vc{y} \in \{0,1 \}^T: \AAS{y} \geq 1}
         \frac{1}{\AAS{y}}
         \left( \sum_{S \in \Sigma_{N,K}} \II( \theta(S) = \vc{y}) \right) \\
    &= \frac{1}{\binom{N}{K}} \sum_{\vc{y} \in \{0,1 \}^T: \AAS{y} \geq 1}
         \frac{1}{\AAS{y}}  \AAS{y}  \\
    &= \frac{| \vc{y} \in \{0,1 \}^T: \AAS{y} \geq 1|}{\binom{N}{K}} 
         \leq \frac{2^T}{\binom{N}{K}},
\end{align*}
since $\{ 0,1 \}^T$, a set of size $2^T$.
\end{IEEEproof}

\begin{figure}
\begin{center}
\includegraphics[width = 3.2in]{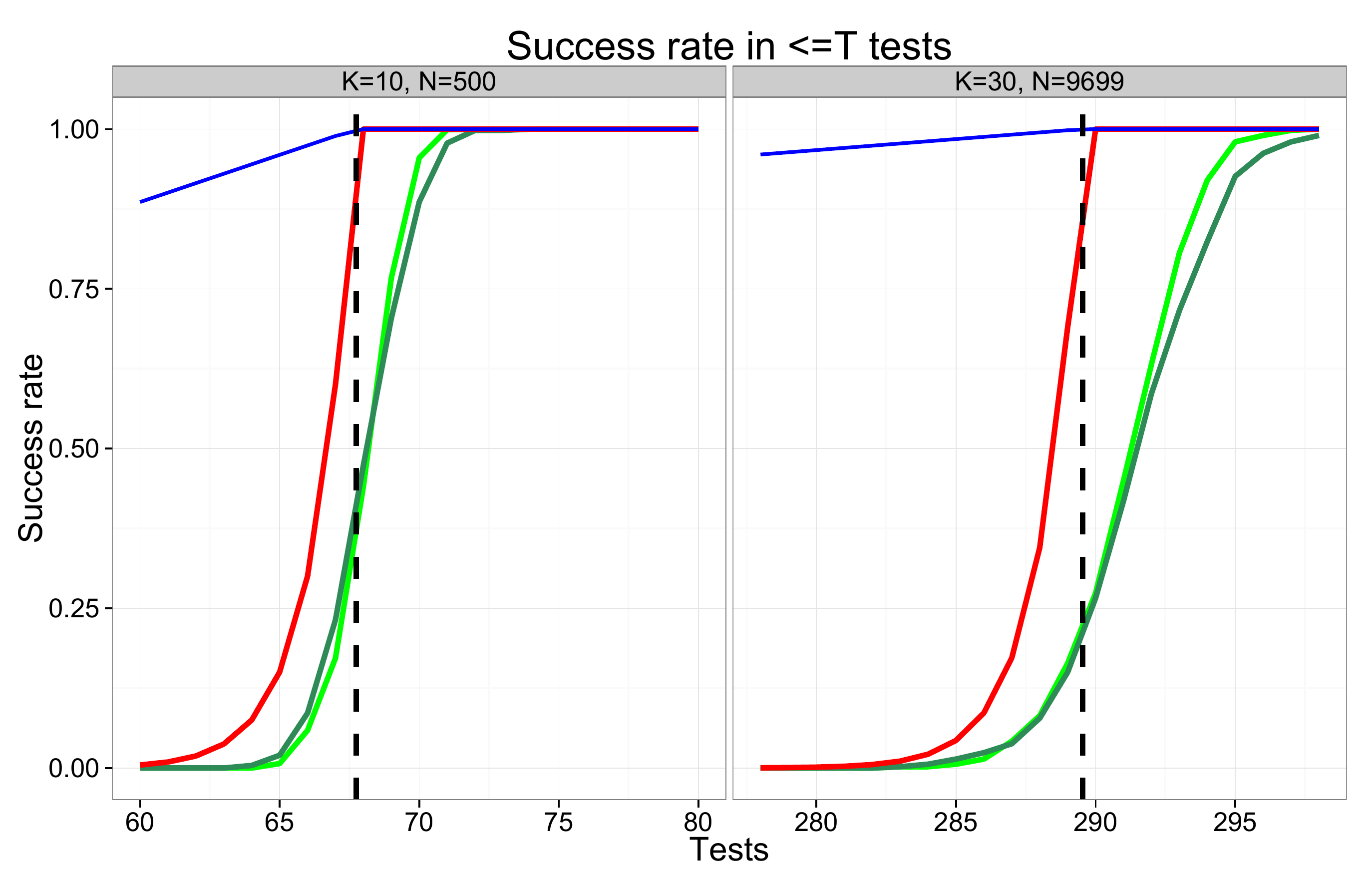}
\caption{%
Plot of success probability of Hwang's group testing algorithms for $(K,N) = (10,500)$ and $(30,9699)$.
 The simulated success probability of the HGBSA is plotted as a bright green line, and the related algorithm analysed in Section \ref{sec:ubound}
is plotted in dark green. The upper bound on
success probability of Theorem
\ref{thm:basic} is plotted in red, and the upper bound of \cite{chan}
Equation (\ref{eq:chan}) in blue. The dotted vertical line
is at $\log_2 \binom{N}{K}$.
 \label{fig:hist}}
 \end{center}
\end{figure}

The fact that $\log_2 \binom{N}{K}$
is the `magic number' of tests providing a lower bound on the number of tests 
required for recovery with success probability 1 is folklore -- see for example 
\cite{hwang}. However, the exponential decay of success probability for lower
numbers of tests which we provide here is new.
Theorem \ref{thm:basic} is a strengthening of Theorem 1 of \cite{chan}, 
which implies that 
\begin{equation}
  \pr( \suc)  \leq \frac{T}{  \log_2 \binom{N}{K}}. \label{eq:chan}
\end{equation}
In fact, Theorem 1 of \cite{chan}
is stated with $K \log_2(N/K)$ in the denominator -- the stronger
form given by (\ref{eq:chan}) is given within their proof, wherein cite{chan} shows that if $\log_2 \binom{N}{K}/T \geq 1+\epsilon$,
  then the success probability is bounded above by $1/(1+\epsilon)$, rather than tending to zero. To 
be precise, Definition \ref{def:capacity} requires a strong converse (in the sense of \cite{wolfowitz}),
whereas \cite{chan} only proves a weak converse.
The differing form of 
(\ref{eq:basic}) and (\ref{eq:chan}) is plotted in Figure \ref{fig:hist},
emphasising that (\ref{eq:basic}) is significantly stronger.

Observe that (using the fact that for any random variable
$\ep T = \sum_{t=0}^\infty (1-\pr(T \leq t))$, 
(\ref{eq:basic}) implies that for any algorithm that uses a random number of tests  $T$
to detect the defective set with certainty,
 the expected success time
\begin{equation} \label{eq:eptbound}
  \ep T \geq \log_2 \binom{N}{K} - 2.
\end{equation}
 We now  prove the main result of the paper, Theorem \ref{thm:main}:

\begin{IEEEproof}[Proof of Theorem \ref{thm:main}]
The result is obtained using the binomial coefficient bounds  (\ref{eq:binomlog}),
with the lower bound meaning that in the regime  $K = o(N)$, then
\begin{equation} \label{eq:binombounds}
  \lim_{N \tends} \frac{\log_2 \binom{N}{K(N)}}{K(N)} =  \infty, 
\end{equation}
since we also have $K \geq 1$.
Now fix $\epsilon > 0$.
First  if, as assumed in (\ref{eq:lower}) for $N$ sufficiently large,
$ T(N) \leq  \frac{1}{1+\epsilon} \log_2 \binom{N}{K(N)}$ then 
Theorem \ref{thm:basic} shows that
$ \pr(\suc) \leq \binom{N}{K(N)}^{-\epsilon/(1+\epsilon)}.$
We deduce the strong converse, i.e. that $\pr(\suc)$ tends to
zero by (\ref{eq:binombounds}).

Theorem \ref{thm:hwang} shows that
Hwang's Generalized Binary Splitting Algorithm is guaranteed to succeed using $T(N) = \log_2 \binom{N}{K(N)} + K(N)$ tests.
We can deduce by 
 (\ref{eq:binombounds}) that $\log_2 \binom{N}{K(N)}/T(N)$ is greater than $1-\epsilon$, for all $N$ sufficiently large,
so (\ref{eq:upper}) follows with $C=1$.
\end{IEEEproof}

\begin{IEEEproof}[Proof of Theorem \ref{thm:noisy}]
Extending this, the capacity of the erasure model for adaptive group testing in the $K=o(N)$ regime is exactly $1-p$.
We simply repeat any erased test until erasure fails to happen and then use Hwang's algorithm. We need to have a number of non-erased tests greater than the bound of Theorem \ref{thm:hwang}. With $\log{N\choose K}/(1-p-\varepsilon)$ tests the probability that this happens approaches 1 exponentially fast. 
\end{IEEEproof}

\section{A tighter upper bound} \label{sec:ubound}

We now give a tighter upper bound on the performance of the HGBSA by controlling the size of the samples to be binary-searched, aiming to make the probability of observing a negative test slightly less than 1/2. This algorithm offers performance similar to the HGBSA, but reduces the gap between the bounds of Theorem \ref{thm:hwang} and Theorem \ref{thm:basic} from $K$ to under $K/2$.

We briefly illustrate the performance of these algorithms by simulation. In the two cases of Figure \ref{fig:hist} we keep the parameter $\beta = 0.63$ fixed and plot the success
probability of the algorithms compared with the lower bound of Theorem
\ref{thm:basic} for different problem sizes.

We now describe the algorithm to prove Theorem \ref{thm:ubound}. We group the tests in $K$  rounds, each of which comprises a sequence of negative tests before a positive test. Once a 
positive test occurs, we can find a defective using binary search as in Lemma \ref{lem:binsearch}.
 We introduce the following notation, which allows us to keep track of the size of each subproblem (round).

\begin{definition}
We refer to an item
which has not yet appeared in a negative test as a Possible Defective. We write $N^{(i)}_j$ the number of Possible Defectives left after the $j$-th consecutive negative test of the $i$-th round
and $N^{(i)}_0$ for the number of Possible Defectives at the start of the $i$-th round, and write $K^{(i)} =
K - i +1$ for the number of defectives in the $i$-th round. We write $\idx Sij$ for the indicator that the 
$j$-th test of the $i$-th round is positive.
Finally define $T_i$ as the random
time of the last negative test of the round.
\end{definition}

After every negative test the set of Possible Defectives will be reduced. In particular, if we denote by $\idx bij$ the size
of the $j$-th test in the $i$-th round, then
\begin{equation} \label{eq:droppd}
  \idx{N}{i}{j+1}=\idx Nij - \idx bi{j+1},\ 1\leq i\leq K,\ 0\leq j\leq T_i\ .
\end{equation}
As described previously, we use  maximally informative tests by
making the probability of observing a negative test (just) less than $1/2$. By truncating binomial coefficients, we prove:

\begin{lemma}
  Conditional on having observed $j-1$ consecutive negative tests in the $i$-th round of Hwang's algorithm, any random sample of size 
    \begin{equation} \label{eq:bchoice}
      \idx bij \leq \idx Ni{j-1} \left(1-2^{-1/\idx Ki{}} \right)-(\idx Ki{} -1)\ .	
    \end{equation}
  has probability less than $1/2$ of being negative, that is
    $$ \pr(\idx Sij=0\given\idx Si0=0, \ldots, \idx {S}{i}{j-1}=0)  \leq 1/2.$$
\end{lemma}

If the test is negative, we update $\idx Nij$ using (\ref{eq:droppd}).
If the test is positive, the only upper bound we can derive for sure for the new size of
the set of possible defectives is $\idx Nij\leq\idx Ni{j-1}-1$, as we're only sure that
one (defective) item will be removed from the set.
By induction, within each round the following formulae for \idx bij and \idx Nij hold:
\begin{eqnarray}
  \idx bij=2^{-\frac{j-1}{K^{(i)}}} \left[N(1-2^{-1/K^{(i)}})-(K^{(i)}-1)\right]\\
  \idx Nij=2^{-\frac{j}{K^{(i)}}}\idx Ni0 + (K^{(i)}-1)\sum_{h=0}^{j-1}2^{-\frac{h}{K^{(i)}}}\ .
\end{eqnarray}
Here \idx Ni0 clearly depends on the previous round; it equals $N$ for $i=1$ and it can be suitably upper bounded for $i\geq2$. With these results we can produce an upper bound on the average number of
tests for this version of Hwang's algorithm.

\begin{theorem}  \label{thm:ubound}
  Our version of Hwang's algorithm is guaranteed to succeed with number of tests satisfying:
\begin{equation}
T_{tot} \leq K\log N + (1+\log\ln2) K - \log K! +R\ .	
\end{equation}
where $R$ is a negative random term.
\end{theorem}
\begin{IEEEproof}
We observe that
\begin{equation} \label{eq:bijubound}
  \idx bij  =  \left\lceil\idx Ni{j-1} \left(1-2^{-1/\idx Ki{}} \right)\right\rceil	\leq  2 \idx Ni{j-1} \left(1-2^{-1/\idx Ki{}} \right)\ .
\end{equation}

Similarly, repeatedly substituting (\ref{eq:bchoice}) in (\ref{eq:droppd}) we obtain
\begin{equation}
  \idx Nij\leq \idx Ni0 2^{-j/\idx Ki{}}\ .
\end{equation}

This can be plugged back into \eqref{eq:bijubound}, giving 
\begin{equation}
	\idx bij \leq 2\idx Ni0 2^{-(j-1)/\idx Ki{}} (1-2^{-1/\idx Ki{}})\ .
\end{equation}
We can then compute the upper bound on the total number of tests using Lemma \ref{lem:binsearch}. The number of
tests in the $i$th round satisfies:
\begin{align}
	T_{tot}^{(i)}	= 	T_i + \log \idx bi{T_i+1} \leq &T_i+ 1 	+	 \log \idx Ni0 \nonumber \\
	& - \df{T_i}{\idx Ki{}} + \log (1-2^{-1/\idx Ki{}}) \ . \label{eq:toti}
\end{align}
In order to obtain a bound on the total number of tests we have to relate $\idx Ni0$ with $\idx N{i-1}{T_{i-1}}$. Recalling Lemma \ref{lem:binsearch}, we define $L_i$ by making $1+L_i$ be the
position of the leftmost defective in the $(T_i+1)$-th sample,
thus obtaining the equality
\begin{equation}
\label{ni0:precise}
	\idx N{i+1}0=\idx Ni{T_i} - L_i\ .
\end{equation}
 Using iteratively the update formulae $\idx Nij\leq \idx Ni0 2^{-j/\idx Ki{}}$ and \eqref{ni0:precise}, we can deduce that $\log \idx N{i+1}0$ equals
 \begin{equation*}
	 \log N - \sum_{j=1}^i \df{T_j}{\idx Kj{}} + \log \left[ 1 - \df{1}{N} \sum_{h=1}^i \df{L_h} {\prod_{j=1}^h 2^{-T_j/\idx Kj{}}} \right]\ .
\end{equation*}
Summing together the bounds for the number of tests in each round, calling $C_i$ the last addend above and $R=\sum_{i=1}^n C_i$:
\begin{align}
	T_{tot}   &\leq 	K\log N + K + \sum_{i=1}^K\log(1-2^{-1/\idx Ki{}}) + R .
			\label{eq:tottests}
\end{align}
To gain a more manageable expression, we  bound  the penultimate term of (\ref{eq:tottests}).
Calling $f(x) := 2^{-x} - 1 + x \ln 2$, notice that $f'(x)\geq0$ for $x\geq0$ and $f(0)=0$; in particular, $1-2^{-1/i}\leq \df{1}{i}\ln2$. Substituting $\idx Ki{}=K-i+1$, reversing the order of the sum and taking logs, we deduce the result. 
\end{IEEEproof}

It is hard to say more about the random terms $C_i$ than that they are negative, though we hope that future
simulation and probabilistic bounds will give us further insights into the resulting number of tests required. Notice the slightly suprising feature that the number of negative 
tests $T_i$ has no effect on the final bound  (\ref{eq:tottests}). This can be explained by the fact that summing (\ref{eq:toti}) over $i$ creates a double sum over $i$ and $j$,
and that the coefficients of $T_i$ exactly cancel.

\section{Conclusion}

Using a sharper
information-theoretic lower bound, we have shown that in the noiseless adaptive case, the capacity
of group testing is 1, and that for  an erasure channel, the capacity is $1-p$. For other noise models, we have
found lower bounds on the capacity.
 It remains of
interest to find exact values of the capacity in other cases, including  non-adaptive problems.

\section*{Acknowledgment}

The authors would like to thank Dino Sejdinovic for useful discussions.
Leonardo Baldassini is supported by a University of Bristol Postgraduate 
Studentship. Matthew Aldridge is supported by the
Heilbronn Institute for Mathematical Research.
%
%

\end{document}